\documentclass[twocolumn,prc,aps,showpacs,floatfix,amsmath,amssymb]{revtex4-2}

\usepackage{graphicx,color}
\usepackage{dcolumn}
\usepackage{bm}
\usepackage{caption} 

\usepackage{subfigure}
\usepackage{caption}
\usepackage{subcaption}
\usepackage{hyperref}
\usepackage{amsmath}
\usepackage{float}
\hypersetup{
    colorlinks,
    citecolor=blue,
    filecolor=blue,
    linkcolor=blue,
    urlcolor=blue
}

\usepackage[bottom]{footmisc}

\newcommand{\sqrtsNN}{\mbox{$\sqrt{\mathrm{s}_{_{\mathrm{NN}}}}$}}

\begin{document}

\title{Comment on ``Evaluation of kinetic freeze-out properties in different relativistic heavy-ion collision systems at \sqrtsNN = 200 GeV'' \\(Eur. Phys. J. Plus (2025) 140:179)\\https://doi.org/10.1140/epjp/s13360-025-06119-0}


\author{M.~U.~Ashraf}
 \email{muashraf@wayne.edu;}
\affiliation{%
Department of Physics and Astronomy, Wayne State University, 666 W. Hancock, Detroit, Michigan 48201, USA\\}

\date{\today}

\begin{abstract}


\end{abstract}

\maketitle


The paper presents transverse momentum spectra for $Zr+Zr$, $Ru+Ru$, and $U+U$ collisions as experimental results (Figs. 2, 3 and 5), citing Ref.~\cite{Sinha:2023jas}. However, Ref.~\cite{Sinha:2023jas} explicitly states these data are ``AMPT model simulations'', not experimental measurements. The study mixes ``experimental data ($Cu+Cu$, $Au+Au$)'' with AMPT-simulated data ($Zr+Zr$, $Ru+Ru$, and $U+U$ ) to draw conclusions about system-size effects.

\begin{itemize}
    \item In the introduction section, the authors inaccurately attribute a misleading claim to the STAR paper. The authors made the following statement ``\textit{In another work~\cite{STAR:2017sal} by the STAR Collaboration, the phase transition has been observed, where up to \(\sqrt{s_{NN}} = 19.6\) GeV, linear growth is reported with collision energy, after which they show a subsequent kink structure.} The STAR Collaboration never claims a ``phase transition'' at \(\sqrt{s_{NN}} = 19.6\) GeV. 
    The ``horn'' in the energy dependence of $K^+/\pi+$ has been suggested as the signature of a phase transition~\cite{STAR:2017sal}. 

\end{itemize}

\section{Kinetic Freeze-out Temperature}

In Section 2, authors correctly define \( T_0 \) as the kinetic freeze-out temperature. However, later, they conflate this with the critical temperature (\( T_c \)) of the QCD phase transition, asserting system-dependent variations in \( T_c \) based on \( T_0 \) trends which is misleading. The authors’ claim that \( T_c \) varies across systems (e.g., $Cu+Cu$ vs. $U+U$) is also not right. Lattice QCD calculations show that \( T_c \) is a universal property of QCD matter, not collision systems~\cite{HotQCD:2014kol}. Further details can be found in Ref.~\cite{STAR:2017sal}.

The authors’ approach of fitting experimental $p_T$ spectra from $Cu+Cu$ and $Au+Au$ collisions while using AMPT-generated data for $Zr+Zr$, $Ru+Ru$, and $U+U$ systems to claim system-dependent critical temperatures ($T_c$) is scientifically unsound. There are various reasons but here are few important:

\begin{itemize}
    \item AMPT is a transport model, lacks first-principles QGP dynamics. Its string-melting mode employs phenomenological hadronization criteria that cannot replicate the QCD phase transition~\cite{Lin:2001yd}.
    \item AMPT is known to poorly describe baryon production at RHIC energies, particularly the $p_T$ spectra of protons and other baryons~\cite{Nandi:2019ztz}. This deficiency directly undermines the reliability of the extracted freeze-out parameters ($T_0$, $\beta_T$), as baryon spectra are critical for constraining kinetic freeze-out conditions.
    \item The authors erroneously equate trends in the kinetic freeze-out temperature ($T_0$)—a late-stage observable—with the critical temperature ($T_c$) of the QGP phase transition, which is an early-stage property governed by QCD thermodynamics. 
    The observed variations in $T_0$ (even if valid) cannot be interpreted as evidence for system-dependent $T_c$.
\end{itemize}

A significant methodological inconsistency undermines the reliability of this analysis: while the original data from Ref.~\cite{Sinha:2023jas} were published without visible error bars (suggesting uncertainties smaller than marker sizes), the current paper inexplicably presents the same data with conspicuously large error bars in its figures. The analysis further suffers from an unacceptable lack of clarity regarding the treatment of nuclear deformation effects in the $Zr+Zr$, $Ru+Ru$, and $U+U$ systems. While in Ref.~\cite{Sinha:2023jas} explicitly investigated multiple deformation scenarios and demonstrated their significant impact on flow observables, the current paper fails to specify which case was adopted for their analysis. This omission is particularly egregious given that nuclear deformation can alter the initial state geometry by up to 20\% in these systems, directly affecting the extracted kinetic freeze-out parameters.

This discrepancy raises serious concerns about proper error propagation and statistical treatment. The apparent inflation of uncertainties in the current work could artificially improve the reported $\chi^2/ndf$ values by making poor fits appear statistically acceptable. For instance, the proton spectra in Fig. 1(c) show substantial deviations from the fitted curve that should normally yield high $\chi^2$ values, yet the claimed $\chi^2/ndf \approx$ 1 becomes plausible only if the uncertainties have been substantially overestimated. Without transparent justification for these modified uncertainties and demonstration that they follow standard error propagation methods, the extracted freeze-out parameters cannot be considered trustworthy. The authors must clarify this discrepancy by either providing rigorous derivation of their error bars or reverting to the original uncertainty representation from Ref.~\cite{Sinha:2023jas}, as the current presentation risks misleading readers about the true precision of the results.



The fitting methodology and parameter interpretation in Figures 6-7 present fundamental inconsistencies that undermine the study's conclusions. While performing independent fits to $\pi^+$, $K^+$ and $p$ spectra, the authors paradoxically claim identical kinetic freeze-out temperatures ($T_0$) and flow velocities ($\beta_T$) across species while allowing the non-extensive parameter ($n$) to vary. This approach is physically unjustified because: Hydrodynamic principles and experimental evidence consistently show mass-dependent freeze-out, where heavier particles (e.g., protons) decouple earlier at higher $T_0$ and lower $\beta_T$ than pions~\cite{Schnedermann:1993ws}. If $T_0$ and lower $\beta_T$ are truly common parameters as asserted, they should be extracted via simultaneous multi-species fits with properly correlated uncertainties~\cite{Andronic:2017pug}. The allowed variation in $n$ contradicts the assumption of a unified freeze-out scenario, as equilibration (encoded in $n$) must be consistently linked to $T_0$ and lower $\beta_T$ through the collision dynamics. The current treatment of parameters—forcing identity for some variables while permitting others to vary freely—reflects either conceptual confusion in the model implementation or selective parameter tuning to achieve desired trends. A rigorous reanalysis using constrained simultaneous fits with proper uncertainty propagation is essential to validate these claims.

\bibliography{bib.bib}

\begin{thebibliography}{7}%
\makeatletter
\providecommand \@ifxundefined [1]{%
 \@ifx{#1\undefined}
}%
\providecommand \@ifnum [1]{%
 \ifnum #1\expandafter \@firstoftwo
 \else \expandafter \@secondoftwo
 \fi
}%
\providecommand \@ifx [1]{%
 \ifx #1\expandafter \@firstoftwo
 \else \expandafter \@secondoftwo
 \fi
}%
\providecommand \natexlab [1]{#1}%
\providecommand \enquote  [1]{``#1''}%
\providecommand \bibnamefont  [1]{#1}%
\providecommand \bibfnamefont [1]{#1}%
\providecommand \citenamefont [1]{#1}%
\providecommand \href@noop [0]{\@secondoftwo}%
\providecommand \href [0]{\begingroup \@sanitize@url \@href}%
\providecommand \@href[1]{\@@startlink{#1}\@@href}%
\providecommand \@@href[1]{\endgroup#1\@@endlink}%
\providecommand \@sanitize@url [0]{\catcode `\\12\catcode `\$12\catcode `\&12\catcode `\#12\catcode `\^12\catcode `\_12\catcode `\%12\relax}%
\providecommand \@@startlink[1]{}%
\providecommand \@@endlink[0]{}%
\providecommand \url  [0]{\begingroup\@sanitize@url \@url }%
\providecommand \@url [1]{\endgroup\@href {#1}{\urlprefix }}%
\providecommand \urlprefix  [0]{URL }%
\providecommand \Eprint [0]{\href }%
\providecommand \doibase [0]{https://doi.org/}%
\providecommand \selectlanguage [0]{\@gobble}%
\providecommand \bibinfo  [0]{\@secondoftwo}%
\providecommand \bibfield  [0]{\@secondoftwo}%
\providecommand \translation [1]{[#1]}%
\providecommand \BibitemOpen [0]{}%
\providecommand \bibitemStop [0]{}%
\providecommand \bibitemNoStop [0]{.\EOS\space}%
\providecommand \EOS [0]{\spacefactor3000\relax}%
\providecommand \BibitemShut  [1]{\csname bibitem#1\endcsname}%
\let\auto@bib@innerbib\@empty
\bibitem [{\citenamefont {Sinha}\ \emph {et~al.}(2023)\citenamefont {Sinha}, \citenamefont {Bairathi}, \citenamefont {Gopal}, \citenamefont {Jena},\ and\ \citenamefont {Kabana}}]{Sinha:2023jas}%
  \BibitemOpen
  \bibfield  {author} {\bibinfo {author} {\bibfnamefont {P.}~\bibnamefont {Sinha}}, \bibinfo {author} {\bibfnamefont {V.}~\bibnamefont {Bairathi}}, \bibinfo {author} {\bibfnamefont {K.}~\bibnamefont {Gopal}}, \bibinfo {author} {\bibfnamefont {C.}~\bibnamefont {Jena}},\ and\ \bibinfo {author} {\bibfnamefont {S.}~\bibnamefont {Kabana}},\ }\href {https://doi.org/10.1103/PhysRevC.108.024911} {\bibfield  {journal} {\bibinfo  {journal} {Phys. Rev. C}\ }\textbf {\bibinfo {volume} {108}},\ \bibinfo {pages} {024911} (\bibinfo {year} {2023})},\ \Eprint {https://arxiv.org/abs/2305.13950} {arXiv:2305.13950 [hep-ph]} \BibitemShut {NoStop}%
\bibitem [{\citenamefont {Adamczyk}\ \emph {et~al.}(2017)\citenamefont {Adamczyk} \emph {et~al.}}]{STAR:2017sal}%
  \BibitemOpen
  \bibfield  {author} {\bibinfo {author} {\bibfnamefont {L.}~\bibnamefont {Adamczyk}} \emph {et~al.} (\bibinfo {collaboration} {STAR}),\ }\href {https://doi.org/10.1103/PhysRevC.96.044904} {\bibfield  {journal} {\bibinfo  {journal} {Phys. Rev. C}\ }\textbf {\bibinfo {volume} {96}},\ \bibinfo {pages} {044904} (\bibinfo {year} {2017})},\ \Eprint {https://arxiv.org/abs/1701.07065} {arXiv:1701.07065 [nucl-ex]} \BibitemShut {NoStop}%
\bibitem [{\citenamefont {Bazavov}\ \emph {et~al.}(2014)\citenamefont {Bazavov} \emph {et~al.}}]{HotQCD:2014kol}%
  \BibitemOpen
  \bibfield  {author} {\bibinfo {author} {\bibfnamefont {A.}~\bibnamefont {Bazavov}} \emph {et~al.} (\bibinfo {collaboration} {HotQCD}),\ }\href {https://doi.org/10.1103/PhysRevD.90.094503} {\bibfield  {journal} {\bibinfo  {journal} {Phys. Rev. D}\ }\textbf {\bibinfo {volume} {90}},\ \bibinfo {pages} {094503} (\bibinfo {year} {2014})},\ \Eprint {https://arxiv.org/abs/1407.6387} {arXiv:1407.6387 [hep-lat]} \BibitemShut {NoStop}%
\bibitem [{\citenamefont {Lin}\ \emph {et~al.}(2002)\citenamefont {Lin}, \citenamefont {Pal}, \citenamefont {Ko}, \citenamefont {Li},\ and\ \citenamefont {Zhang}}]{Lin:2001yd}%
  \BibitemOpen
  \bibfield  {author} {\bibinfo {author} {\bibfnamefont {Z.-w.}\ \bibnamefont {Lin}}, \bibinfo {author} {\bibfnamefont {S.}~\bibnamefont {Pal}}, \bibinfo {author} {\bibfnamefont {C.~M.}\ \bibnamefont {Ko}}, \bibinfo {author} {\bibfnamefont {B.-A.}\ \bibnamefont {Li}},\ and\ \bibinfo {author} {\bibfnamefont {B.}~\bibnamefont {Zhang}},\ }\href {https://doi.org/10.1016/S0375-9474(01)01383-5} {\bibfield  {journal} {\bibinfo  {journal} {Nucl. Phys. A}\ }\textbf {\bibinfo {volume} {698}},\ \bibinfo {pages} {375} (\bibinfo {year} {2002})},\ \Eprint {https://arxiv.org/abs/nucl-th/0105044} {arXiv:nucl-th/0105044} \BibitemShut {NoStop}%
\bibitem [{\citenamefont {Nandi}\ \emph {et~al.}(2020)\citenamefont {Nandi}, \citenamefont {Kumar},\ and\ \citenamefont {Sharma}}]{Nandi:2019ztz}%
  \BibitemOpen
  \bibfield  {author} {\bibinfo {author} {\bibfnamefont {A.}~\bibnamefont {Nandi}}, \bibinfo {author} {\bibfnamefont {L.}~\bibnamefont {Kumar}},\ and\ \bibinfo {author} {\bibfnamefont {N.}~\bibnamefont {Sharma}},\ }\href {https://doi.org/10.1103/PhysRevC.102.024902} {\bibfield  {journal} {\bibinfo  {journal} {Phys. Rev. C}\ }\textbf {\bibinfo {volume} {102}},\ \bibinfo {pages} {024902} (\bibinfo {year} {2020})},\ \Eprint {https://arxiv.org/abs/1910.11558} {arXiv:1910.11558 [hep-ph]} \BibitemShut {NoStop}%
\bibitem [{\citenamefont {Schnedermann}\ \emph {et~al.}(1993)\citenamefont {Schnedermann}, \citenamefont {Sollfrank},\ and\ \citenamefont {Heinz}}]{Schnedermann:1993ws}%
  \BibitemOpen
  \bibfield  {author} {\bibinfo {author} {\bibfnamefont {E.}~\bibnamefont {Schnedermann}}, \bibinfo {author} {\bibfnamefont {J.}~\bibnamefont {Sollfrank}},\ and\ \bibinfo {author} {\bibfnamefont {U.~W.}\ \bibnamefont {Heinz}},\ }\href {https://doi.org/10.1103/PhysRevC.48.2462} {\bibfield  {journal} {\bibinfo  {journal} {Phys. Rev. C}\ }\textbf {\bibinfo {volume} {48}},\ \bibinfo {pages} {2462} (\bibinfo {year} {1993})},\ \Eprint {https://arxiv.org/abs/nucl-th/9307020} {arXiv:nucl-th/9307020} \BibitemShut {NoStop}%
\bibitem [{\citenamefont {Andronic}\ \emph {et~al.}(2018)\citenamefont {Andronic}, \citenamefont {Braun-Munzinger}, \citenamefont {Redlich},\ and\ \citenamefont {Stachel}}]{Andronic:2017pug}%
  \BibitemOpen
  \bibfield  {author} {\bibinfo {author} {\bibfnamefont {A.}~\bibnamefont {Andronic}}, \bibinfo {author} {\bibfnamefont {P.}~\bibnamefont {Braun-Munzinger}}, \bibinfo {author} {\bibfnamefont {K.}~\bibnamefont {Redlich}},\ and\ \bibinfo {author} {\bibfnamefont {J.}~\bibnamefont {Stachel}},\ }\href {https://doi.org/10.1038/s41586-018-0491-6} {\bibfield  {journal} {\bibinfo  {journal} {Nature}\ }\textbf {\bibinfo {volume} {561}},\ \bibinfo {pages} {321} (\bibinfo {year} {2018})},\ \Eprint {https://arxiv.org/abs/1710.09425} {arXiv:1710.09425 [nucl-th]} \BibitemShut {NoStop}%
\end{thebibliography}%


\end{document}